\documentclass[twocolumn]{aastex61}

\usepackage{longtable}
\usepackage{graphicx}
\usepackage{ulem}
\usepackage{verbatim}

\received{June 28, 2018}
\revised{July 21, 2018}
\accepted{July 26, 2018}
\submitjournal{ApJL}

\shorttitle{A Submillimeter Burst of S255}
\shortauthors{Liu, Su, Zinchenko, et al.}

\begin{document}

\title{A Submillimeter Burst of S255IR~SMA1 - The Rise And Fall Of Its Luminosity}

\correspondingauthor{Sheng-Yuan Liu}
\email{syliu@asiaa.sinica.edu.tw}

\author{Sheng-Yuan Liu}
\affil{Institute of Astronomy and Astrophysics, Academia Sinica, 11F of ASMAB, AS/NTU No.1, Sec. 4, Roosevelt Road, Taipei 10617, Taiwan}

\author{Yu-Nung Su}
\affil{Institute of Astronomy and Astrophysics, Academia Sinica, 11F of ASMAB, AS/NTU No.1, Sec. 4, Roosevelt Road, Taipei 10617, Taiwan}

\author{Igor Zinchenko}
\affil{Institute of Applied Physics of the Russian Academy of Sciences, 46 Ul'yanov str., 603950, Nizhny Novgorod, Russia}

\author{Kuo-Song Wang}
\affil{Institute of Astronomy and Astrophysics, Academia Sinica, 11F of ASMAB, AS/NTU No.1, Sec. 4, Roosevelt Road, Taipei 10617, Taiwan}

\author{Yuan Wang}
\affil{Max Planck Institute for Astronomy, K\"onigstuhl 17, D-69117, Heidelberg, Germany}

\begin{abstract}

Temporal photometric variations at near infrared to submillimeter wavelengths have been found in low-mass young stellar objects.
These phenomena are generally interpreted as accretion events of star-disk systems with varying accretion rates.
There is growing evidence suggesting that similar luminosity flaring also occurs in high-mass star/cluster-forming regions.
We report in this Letter the rise and fall of the 900 ${\mu}$m continuum emission and the newly found 349.1 GHz methanol maser emission in the massive star forming region S255IR~SMA1 observed with the Submillimeter Array and the Atacama Large Millimeter/submillimeter Array. The level of flux variation at a factor of $\sim$ 2 at the submillimeter band and the relatively short 2-year duration of this burst suggest that the event is probably similar to those milder and more frequent minor bursts seen in 3D numerical simulations. 

\end{abstract}


\section{Introduction} \label{sec:intro}

Temporal photometric variations associated with young low-mass stars of Class I or II, such as those FU Orion-type (FUor) and EX Lup-type (EXor) events, have been observed in the optical to mid-infrared (MIR)  wavelengths \citep{Audard14}.
Accompanied by spectroscopic signatures of hot disks and winds, these phenomena are generally interpreted as accretion events of star-disk systems with elevated rates ---
instead of falling through steady flows to the central stars, circumstellar material fragments and accretes sporadically due to instabilities developed in the disks.
Supplementary evidence includes the semi-periodic ejection events seen in the Herbig-Haro (HH) objects associated with low-mass young stellar objects (YSOs).

It is conceivable that at an earlier, more embedded (Class 0/I) evolutionary stage, YSOs may subjected to similar episodic accretion events. 
Due to envelope obscuration, however, such phenomenon may not be visible in the optical or infrared (IR) but possibly detectable at longer (far-infrared (FIR) to millimeter) wavelengths \citep{Johnstone13}.
Indeed, HOPS~383, a Class 0 protostar, was the very first example reported with a brightening event not only in the MIR but also in the submillimeter bands \citep{Safron15}.
Furthermore, recent submillimeter observations of YSOs in nearby molecular clouds successfully revealed for the first time through a monitoring program a luminosity flaring event toward a Class I YSO in the Serpens cloud \citep{Yoo17}.
Molecular line imaging experiments, probing the thermal history of envelopes around embedded YSOs, also provided indirect indications of luminosity flarings \citep{Jorgensen13}.

Does a similar luminosity variation phenomenon occur in the massive star formation process? 
There now appears to be growing evidence pointing to a positive answer. 
For example, the massive star-cluster-forming region S255 was first detected with methanol maser flares \citep{Fujisawa15}.
Subsequent observations in the near infrared (NIR) witnessed brightening of not only the NIR continuum but also atomic and molecular lines \citep{Caratti17}. 
NGC 6334(I), another massive star-cluster-forming site, also got spotted with an increase of its submillimeter continuum \citep{Hunter17}.
Follow-up observations confirm the emergence of new methanol masers 
as a result of the luminosity burst \citep{Hunter18}.

S255IR is a massive cluster formation region with a bolometric luminosity of several 10$^{4}$ L$_\odot$ at a distance of 1.78$^{+0.12}_{-0.11}$ kpc \citep{Burns16}. 
NIR imaging revealed a cluster of YSOs associated with the molecular gas ridge sandwiched by two nearby H{\small II} regions \citep{Ojha11}.
Submillimeter continuum observations indicated an overall molecular gas of around 300---400 M$_\odot$ in this region, and disclosed at higher angular resolution several dense clumps residing in the complex
\citep{Wang11, Zinchenko12, Zinchenko15}.
In particular, the dominant source, S255IR~SMA1, coinciding with the NIR source S255IR~NIRS3, is associated with a prominent molecular bipolar outflow and a rotating disk-like structure \citep{Zinchenko12, Zinchenko15}. 
The putative disk, perpendicular to the outflow in its orientation, probably is viewed closely edge-on \citep{Boley13}.
Based on its luminosity and the maser kinematics, the mass of the central YSO is estimated to be $\sim$ 20~M$_\odot$ \citep{Zinchenko15}.
This region gained appreciable attention for its flaring signatures in 6.7 GHz methanol maser, NIR, and radio continuum \citep{Fujisawa15, Stecklum16, Caratti17, Cesaroni18}
Meanwhile, \cite{Zinchenko17} reported a newly identified submillimeter methanol maser, likely associated with this flare event, too.

\section{Observations}

This study utilizes observations with both the Submillimeter Array (SMA) and the Atacama Large Millimeter/submillimeter Array (ALMA). The SMA observations conducted in 2010 had been described in \citet{Zinchenko15}.
The ALMA observations include three epochs, with the first two being introduced by \citet{Zinchenko17}.
The third epoch of ALMA observations was carried out in the Cycle~4 Science Operation under \#2015.1.00500.S.
This last single execution block was conducted on 2017 July 20
when the 12 m array was in its C40-5 configuration 
with 43 antennas online.
The projected baselines range between 15 m and 3.0 km.
The same set of spectral windows in \citet{Zinchenko17} are again employed.
The continuum window is centered at 335.4 GHz with a bandwidth of 1875.0 MHz.
The window at 349.0 GHz, covering the methanol maser, has 
a spectral channel width of 0.244 MHz and a resolution of 0.488 MHz (0.42 km s$^{-1}$) with online Hanning-smooth applied.
J0854+2006 and J0750+1231 were used as the bandpass and flux calibrators, respectively, and J0613+1708 served as the complex gain calibrator.
A compilation of key parameters for all observations are listed in Table~\ref{tab:obssetup}.

The calibration and reduction of ALMA data were carried out with the Common Astronomy Software Applications \citep{McMullin07} and delivered by the observatory.
Further imaging of the continuum emission is achieved by using the visibilities in the continuum window after obvious spectral contamination removed.
Self-calibration were employed separately for different observational epochs as described in \citet{Zinchenko17} for improving the imaging quality and dynamical range.
The antenna gain solutions were applied to both the continuum data as well as the 349.0 GHz spectral window.

\section{Results}
\label{sec:results}

\subsection{Submillimeter dust continuum \label{result-cont}} 

\subsubsection{SMA versus ALMA \label{result-sma-alma}}

\begin{figure*}
\includegraphics[width=18cm]{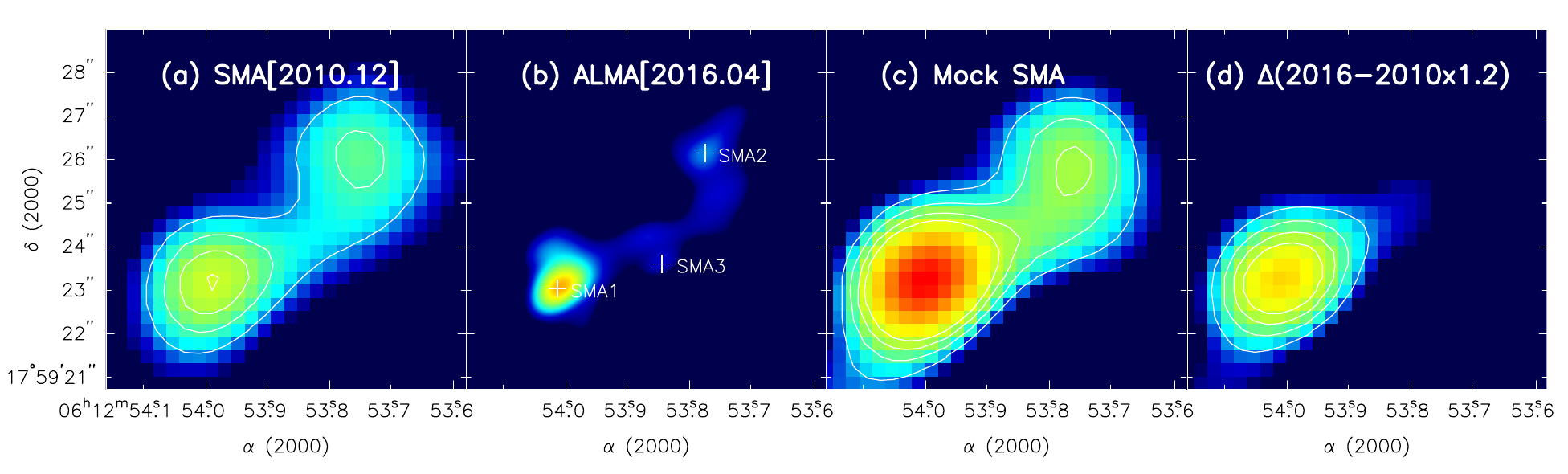}
\caption{
\label{fig:f1}
900 ${\mu}$m continuum image of S255IR (a) observed in 2010 December by SMA at an angular resolution of $\sim$ 2\farcs0. Contour levels are at 3, 5, 7, and 9 $\times$ 46~mJy/beam. 
(b) Observed in 2016 April by ALMA at an angular resolution of $\sim$ 0\farcs6. The positions of SMA1--3 are marked by white crosses.
(c) Made through mock SMA observations using panel (b) as the sky model.
(d) The difference map made by first scaling panel (a) by 1.2 and subtracting that from panel (c). Contour levels in (c) and (d) are the same as (a), so does the false color scheme.
}
\end{figure*}

Panels (a)---(b) in Figure~\ref{fig:f1} present the 900 ${\mu}$m continuum images of S255IR observed with SMA in 2010 December as reported in \cite{Zinchenko15} and with ALMA in 2016 April.
At a lower (SMA) resolution (panel (a)), two prominent features, S255IR~SMA1 and SMA2, are resolved apart.
S255IR~SMA3 is further revealed in the higher resolution ALMA image (panel (b)).
To compare the two observations fairly, we made mock SMA observations by re-sampling and imaging the ALMA map (panel (b)) using the $uv$-coverage achieved by the 2010 SMA observation, similar to the approach adopted in \citet{Hunter17}.
In the resulting Figure~\ref{fig:f1}(c), S255IR~SMA1 and SMA2 remain visible in the mock image.
The SMA image was taken at a slightly higher frequency (343 GHz) while the mock SMA image was taken at a lower frequency (335 GHz).
Considering a typical spectral index of 3 for dust emission, one may expect the SMA image to be 7\% brighter.
However, it shows overall lower intensities.
Instead of owing this to a negative spectral index, the discrepancy likely arises from the absolute flux calibration uncertainties, which can be as much as 20 -- 30\% for the SMA observations in particular.
Furthermore, SMA1 and SMA2 show disparate contrasts in the two epochs, hinting on flux density variation in SMA1 and/or SMA2 during this period.
It is very unlikely that the SMA1 flux remained constant, as this would imply absolute flux calibration errors being greater than 50\%. 
Therefore, displayed in Figure~\ref{fig:f1}(d) is the difference map between panels (a) and (c), if we assume that SMA2 has not changed. For that, (a) was scaled up by 20\%, reasonably within calibration uncertainties, and subtracted from (c).
A compact excess emission feature is visible in this difference map toward SMA1 while no obvious residual remains toward SMA2.

Given prior evidences of a flaring event associated with SMA1 in mid 2015, the above excess emission is most probably associated with this event.
As listed in Table~\ref{tab:obssetup}, the peak intensity in SMA1 rose from 0.50~Jy/beam in 2010 (after scaling) to 1.1~Jy/beam in 2016.
The total flux density of SMA1, meanwhile, picked up from 0.73~Jy in 2010 (also after scaling) to 1.40~Jy in 2016. 
In brief, both its intensity and flux density roughly doubled in 2016 as compared to 2010.

\subsubsection{High-resolution imaging with ALMA \label{result-alma-ext}}

\begin{figure*}
\includegraphics[width=18cm]{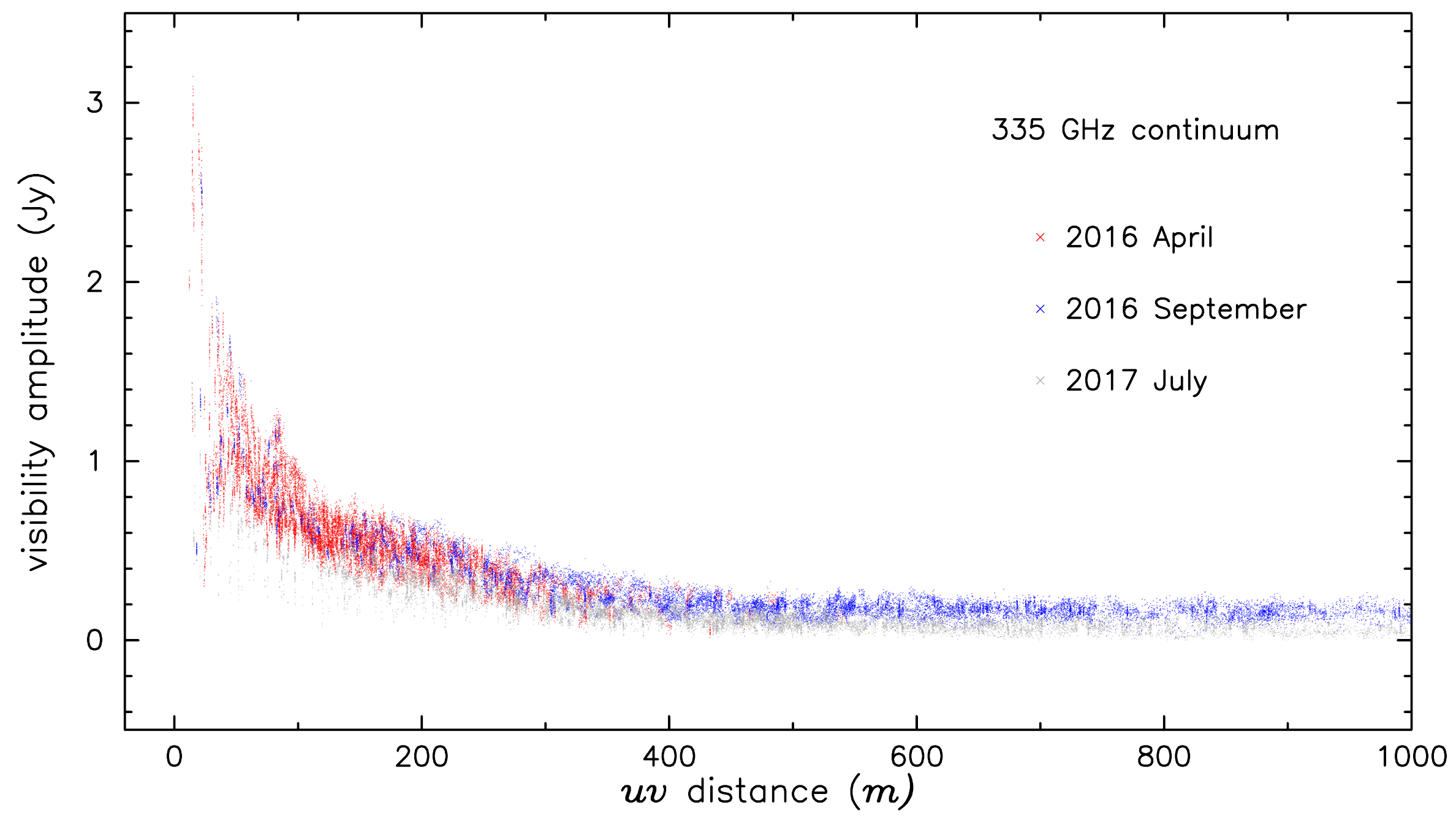}
\caption{
\label{fig:f2}
Visibility amplitudes of the 335 GHz continuum plotted against the $uv$-distance in meters. Red, blue, and gray points represent data taken with ALMA in 2016 April, 2016 September, and 2017 July, respectively.}
\end{figure*}

We present in Figure~\ref{fig:f2} the 900 ${\mu}$m continuum visibilities obtained with ALMA in 2016 April, 2016 September, and 2017 July.
It is evident that between the first two epochs the visibilities at overlapping $uv$ ranges are fully compatible, implying
no variation toward S255IR during this period.
On the other hand, variation (decreasing) of flux is evident in 2017 July.
To localize this variation, we display in Figures~\ref{fig:f3} (a)---(b) the high (0\farcs14) resolution 900 ${\mu}$m continuum observed in 2016 September and 2017 July.
For comparing these two observations, we generated synthesis images by using data within a common $uv$-range (14.5---1765 m) well shared by both observations and then restoring the final resolution to a circular 0\farcs14 beam.
Figure~\ref{fig:f3}(c) is the difference map made by subtracting panel (b) from panel (a).
While emission features such as SMA2 and SMA3 are canceled out, excess emission stands out noticeably at S255IR~SMA1.

The peak intensity in S255IR~SMA1, as observed in these maps, dropped from 0.237~Jy/beam, corresponding to a brightness temperature of 131~K, in 2016 September to 0.137~Jy/beam, or equivalently 76~K, in 2017 July.
The flux density within a 0\farcs5 aperture centered at SMA1 decreased by 33\% during this interval.

\subsection{Methanol maser emission \label{result-maser}}

\cite{Zinchenko17} reported a nonthermal methanol emission line at 349.1 GHz, reaching a brightness temperature of 3900~K (5.9~Jy/beam) at an angular resolution of $\sim$ 0\farcs12. 
The line feature was assigned as a newly discovered maser associated with the CH$_3$OH $14_1 - 14_0 A^{-+}$ transition, likely a Class II CH$_3$OH maser predominantly excited by IR radiation field.
As indicated in \citet{Zinchenko17}, there was no indication of intensity variation in maser emission between 2016 April and September.

In Figure~\ref{fig:f4}, the spectra at its peak position and the integrated intensity of the CH$_3$OH maser observed in 2016 September and 2017 July are presented.
The peak intensities, though remaining nonthermal-like, has discernibly fallen to 2380~K (3.6~Jy/beam) in 2017 July, a reduction of 40\%.
The integrated intensity maps suggested no obvious shifts in the peak location within this period, while the difference map of the integrated intensity (Figure~\ref{fig:f4}(c)) indicates that the excess CH$_3$OH emission region falls within the extent of the (excess) continuum emission delineated by the contour.

\section{Discussion}
\label{sec:discussion}

\subsection{A submillimeter flare, a manifestation of a luminosity burst}

As illustrated in the Section~\ref{sec:results}, the comparison between our SMA and ALMA observations indicated a factor of 2 increase in both the intensity and flux density of 900 ${\mu}$m continuum toward SMA1 in early 2016. ALMA observations further witnessed, for the first time, the waning of this continuum emission as well as CH$_3$OH maser in mid 2017.

No matter whether the dust continuum emission is optically thick or thin, its brightness and flux density variation is reflecting a dust temperature change, unless there is a dramatic modification in dust column density and/or opacity throughout the region.
While an increase in luminosity may lead to enhanced sublimation of dust grains in the close vicinity ($\gtrsim$ 10 au) of the central star and alter the heating timescale, the bulk dust properties likely remain similar \citep{Hunter17}.
The dust temperature elevated by a factor of $\sim$ 2 in early 2016, thus suggests an overall bolometric luminosity increase by a factor of about 16 in S255IR~SMA1 as total emission of the dust envelope scales with temperature to the 4th power \citep{Hunter17}.
For the short cooling timescale (of a few hundred seconds) of dust estimated from its heat capacity \citep{Johnstone13} and cooling rate \citep{Glover12}, 
the 88\% dimming submillimeter continuum in 2016---2017, correspondingly a decreasing of dust temperature by 40\%, also plausibly reflects and a reduced radiation field, as well as a factor of $\sim$ 8 decrease in its luminosity.
Indeed, we witnessed fading (methanol) maser intensity along the same line of sight, most likely due to this reduction in seed radiation.

By investigating the NIR to submillimeter spectral energy distribution (SED) taken at the pre-burst phase and at 2016 February, \cite{Caratti17} unveiled a boost dominantly in the FIR at around 30---100 ${\mu}$m.
The bolometric luminosity of S255IR~SMA1/NIRS3 grew from $3\times 10^4$~L$_\odot$ to $1.6 \times 10^5$~L$_\odot$, a factor of $\sim$ 5.5 increase.
This boost factor appears to be smaller than what we derived.
Given that their FIR data are taken at angular resolutions coarser than 6\farcs0, the inclusion of SMA2 and the surrounding material within those SED measurements probably resulted in lowering the boost factor.
Furthermore, our luminosity estimation has a strong (fourth power) dependence on the dust temperature; a 10\% uncertainty in the flux density, hence temperature measurement could lead to 40\% uncertainty in the derived luminosity, which may partly mitigate the differences in the boost factors.

\subsection{The sequence of events}

Based on the propagation of the light echo, \citet{Caratti17} hypothesized that the S255IR~SMA1 burst event occurred in mid-June of 2015.
This appears consistent with the first report of 6.7 GHz methanol maser flaring seen from early July of 2015 as reported by \citet{Fujisawa15}. 
Direct NIR imaging in 2016 April disclosed brightening of the region as compared to the pre-burst image taken in 2009 \citep{Caratti17}.
The extended $K_s$-band emission is the reprocessed light from hot dust presumably heated by the central star and escaped from the outflow cavity.
Our ALMA image taken in 2016 April, meanwhile, also exhibits boosted submillimeter intensity toward SMA1 as compared to that of 2010.
This emission presumably originates from the dusty disk and/or envelope, which remains optically thick to NIR and processes the radiation to longer wavelengths.
Based on their Karl G. Jansky Very Large Array (VLA) monitoring observations, \citet{Cesaroni18} further reported exponential flux flaring in the radio continuum between 6 and 45 GHz associated with SMA1 from 2016 July to 2017 February.
This rising radio emission was interpreted as the radio jet breakout.
Our 2017 July observation of both dust continuum and methanol maser subsequently indicates the dimming of this burst and marks the burst duration at around two years.

\subsection{Origin of the luminosity burst}

\begin{figure*}
\includegraphics[width=18cm]{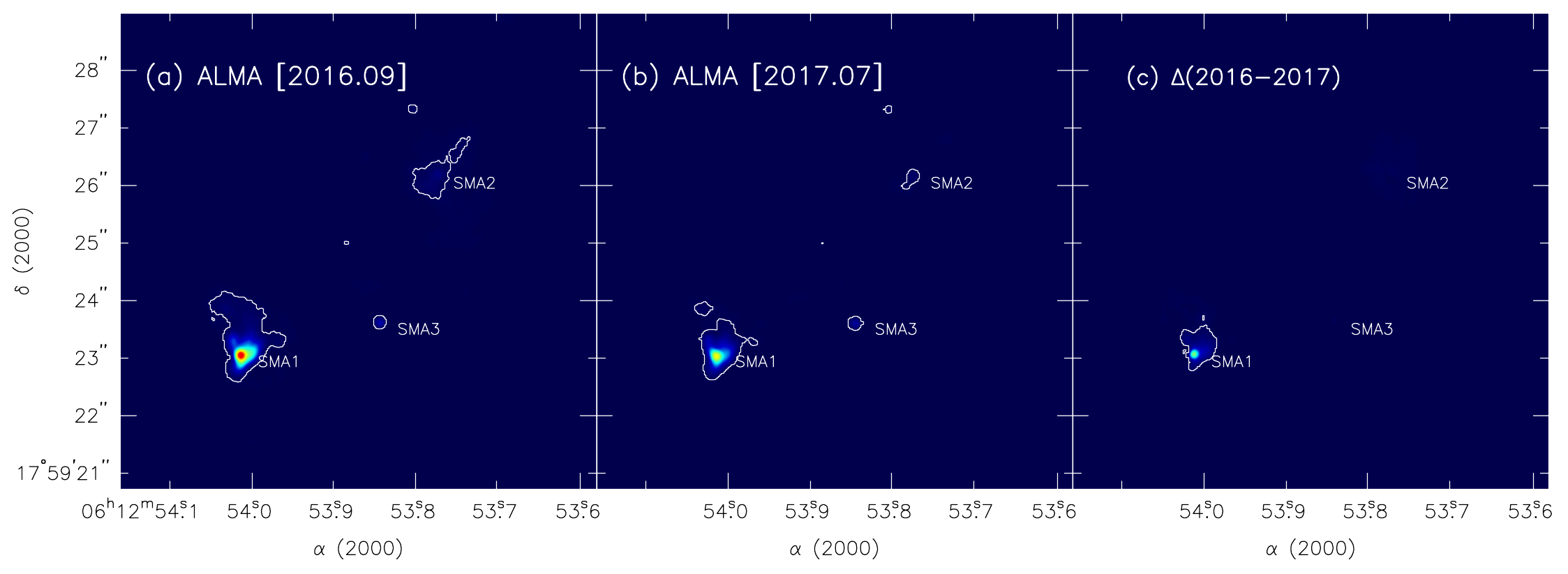}
\caption{
\label{fig:f3}
900 ${\mu}$m continuum image of S255IR (a) observed in 2016 September by ALMA in false color at an angular resolution of 0\farcs14. (b) Observed in 2017 July by ALMA at an angular resolution of 0\farcs14. The contour and labels are the same as those in (a).
(c) The difference map made by subtracting panel (b) from panel (a). The contour at 5-$\sigma$ level marks the boundary of regions with significant emission and SMA1--3 are labeled in panel (a)--(c).
}
\end{figure*}

\begin{figure*}
\includegraphics[width=18cm]{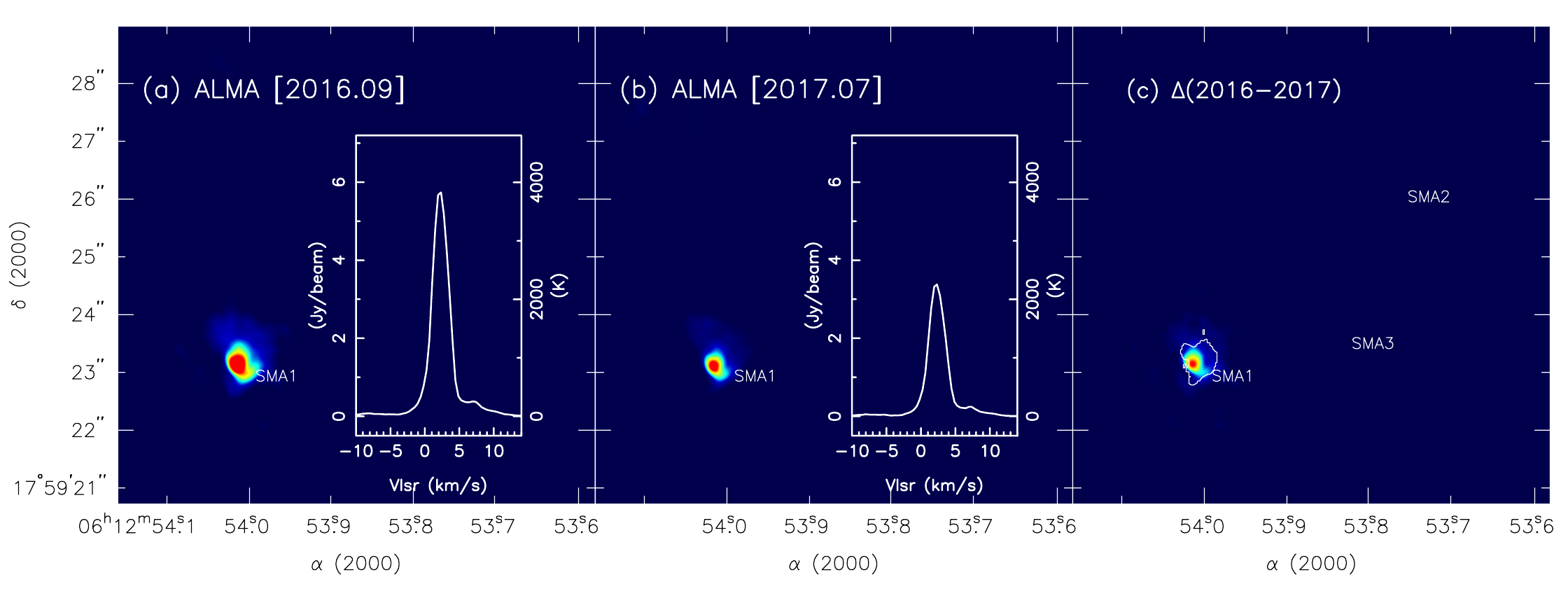}
\caption{
\label{fig:f4}
(a) Integrated intensity map of the 349.1 GHz 
CH$_3$OH $14_1-14_0 A^{-+}$ maser emission observed by ALMA in 2016 September. An inset in the panel displays the CH$_3$OH spectra at its peak position.
(b) The integrated intensity map of the same maser emission observed by ALMA in 2017 July. An inset, same as in panel (a), displays the peak position spectra.
(c) The difference CH$_3$OH maser map made by subtracting panel (b) from (a). The contour delineates the region with excess 900 ${\mu}$m continuum emission shown in Figure~\ref{fig:f3}(c).
}
\end{figure*}

The fact that the submillimeter continuum emission brightened up and dimmed down in conjunction with Class II methanol maser activities in S255IR~SMA1 is suggestive of variation in the IR radiation field and supports the notion of a disk-mediated accretion-related event as suggested by \citet{Caratti17}.
Indeed, on the large (10,000 au) scale, several lines of evidence point to the existence of a rotating structure associated with S255IR~SMA1. 
\citet{Wang11} and \citet{Zinchenko12} demonstrated a velocity gradient perpendicular to the bipolar outflow in several molecular tracers, including CH$_3$CN, HCOOCH$_3$, C$^{18}$O, and CH$_3$OH.
While the angular resolution was not sufficient for a firm conclusion, the kinematic signatures of CH$_3$OH emission, including position-velocity diagram and velocity dispersion map, were both indicative of spin-up toward the inner region, reminiscent of Keplerian rotation \citep{Zinchenko15}.
MIR imaging by \cite{Boley13} resolved an elongated dusty structure on a 1000 au scale perpendicular also to the bipolar outflow, possibly representing the putative circumstellar disk around S255IR~SMA1. 
\cite{Caratti17} further reasoned that the accretion burst is disk-mediated based on the observed emission lines, such as H$_2$, Br$\gamma$, as well as CO band-heads, He \small{I}, and Na \small{I}, typically seen as signatures of enhanced accretion, wind, and hot disks in low-mass YSOs. The compact 900 ${\mu}$m continuum excess region seen in Figure~\ref{fig:f3}(c) is likely associated with this structure.

Variable accretion is not uncommon in 3D numerical radiation hydrodynamic simulations of (low-mass and primordial) star formation \citep[e.g.][]{Vorobyov15,Hosokawa16}.
Asymmetric features like spirals form in the circumstellar disk, which leads to fluctuating accretion rates. 
Moreover, \citet{Meyer17} suggested the universal sporadic and variable nature of gaseous material accreting from the circumstellar disks to massive YSOs as in their low-mass counterparts. 
In their investigations of the collapse of 100 M$_{\odot}$ pre-stellar cores, clumps of molecular gas spiraling from a few hundred au in the disk down to a few tens au from the central massive YSO and episodically accretes onto the star.
In one particular case, the accretion rates and accompanying luminosity have major jumps up by nearly two orders of magnitudes for 10 years scale, while small variations in accretion rate and luminosity take place frequently.
Such a disk fragmentation scenario was invoked as a possible cause leading to the quadruple luminosity burst seen toward NGC 6334(I) \citep{Hunter17}. 

For the low-mass star case, bursts of FUors and EXors are categorized with different characteristics.
While FUor bursts typically arise with increases in mass accretion by several thousand folds and persist for years to decades, EXor bursts have enhanced mass accretion by factors of tens to hundreds and last for just months to years \citep{Hartmann16}.
Considering the relative mild magnitude and the short duration of the flare of S255IR~SMA1,
its temporal behavior resembles the milder and more frequent burst events, in analog to those seen in the EXors for low-mass YSOs.
That being said, in an absolute sense, the scale of energy released by massive YSO events like this one in S255IR~SMA1 or the burst seen in NGC 6334(I) \citep{Hunter17} is dramatically different from the low-mass YSO cases.
The luminosity of S255IR~SMA1 at its burst stage, reaching $10^5 L_{\odot}$, corresponds to a mass accretion rate of several times 10$^{-3}$ M$_\odot$ \citep{Caratti17}. This is far more significant when compared with even the FUor bursts.
 
Based on the multiple scale outflows observed toward S255IR~SMA1, \cite{Burns16} conjectured an episodic accretion history over the last few thousand-year timescale.
The most recent methanol maser and IR flare combined with our ALMA observations revealing the brightening and dimming signatures in submillimeter continuum further pointed out to an ongoing accretion burst.
Monitoring observations of S255IR~SMA1 in its continuum emission shall further constrain the full duration and degree of this latest burst.
Meanwhile, supplementary molecular line observation could gauge the gas temperature, which should be subsequently warmed and cooled along the event, although the expected larger thermal time constant may imply a longer time lag and/or perhaps a milder strength in the gas temperature variation \citep{Johnstone13}.
Finally, high-angular-resolution imaging is essential for resolving the disk structure and revealing the burst nature of S255IR~SMA1.

\begin{deluxetable*}{lcccc}
\tablecaption{Observing Parameters and SMA1 Measurements\label{tab:obssetup}}
\tablewidth{0pt}
\tablehead{
\colhead{Array} & \colhead{SMA} & \multicolumn3c{ALMA} \\
\cline{3-5}
}
\startdata
Observation date & 2010 Dec 15 & 2016 Apr 21 & 2016 Sep 09 & 2017 Jul 20 \\
Configuration & Compact & C36-2/3 & C36(5)/6 [=C40-6] & C40-5 \\
On-source time (minutes)&  135 & 43 & 86 & 43 \\
Antenna number & 8 & 42 & 39 & 43 \\
Minimum projected baseline (m) & 9.1 & 12.1 & 12.0 & 14.6 \\
Maximum projected baseline (m) & 77 & 562 & 2811 & 3041 \\
Continuum (imaging) freq. (GHz)  & 343.0 & 335.4 & 335.4 & 335.4 \\
Continuum bandwidth (GHz) & 2 & 1.875 & 1.875 & 1.875 \\
\cline{1-5}
SMA1 900 ${\mu}$m continuum & & & & \\
peak brightness (Jy beam$^{-1}$) & 0.50\tablenotemark{a} & 1.1\tablenotemark{a} & 0.24\tablenotemark{b} & 0.14\tablenotemark{b} \\
flux density (Jy) & 0.73\tablenotemark{c} & 1.4\tablenotemark{c} & 0.64\tablenotemark{d} & 0.41\tablenotemark{d} \\
\cline{1-5}
SMA1 CH$_3$OH $14_1-14_0 A^{-+}$ maser & & & & \\
peak brightness (Jy/beam) &   & 5.9\tablenotemark{e} & 5.9\tablenotemark{e} & 3.6\tablenotemark{e} \\
\enddata
\tablenotetext{a}{For a 2\farcs3 $\times$ 1\farcs9 beam with a position angle -73$^{\circ}$}
\tablenotetext{b}{For a 0\farcs14 $\times$ 0\farcs14 beam}
\tablenotetext{c}{For a 4\farcs0 aperture}
\tablenotetext{d}{For a 0\farcs5 aperture}
\tablenotetext{e}{For a 0\farcs14 $\times$ 0\farcs14 beam}
\end{deluxetable*}

S.Y.L. acknowledges the support by the Minister of Science and Technology of Taiwan (MOST 106-2119-M-001-013).
I.Z.'s research was supported by the Russian Science Foundation (grant No. 17-12-01256).
This Letter makes use of the following ALMA data: ADS/JAO.ALMA \#2015.1.00500.S. ALMA is a partnership of ESO (representing its member states), NSF (USA) and NINS (Japan), together with NRC (Canada), MoST and ASIAA (Taiwan), and KASI (Republic of Korea), in cooperation with the Republic of Chile. The Joint ALMA Observatory is operated by ESO, AUI/NRAO and NAOJ.

\end{document}